\documentclass[aps, prd, twocolumn, showpacs, superscriptaddress, groupedaddress]{revtex4} 
\usepackage{graphicx}	
\usepackage{amssymb}
\usepackage{dcolumn}
\usepackage[mathscr]{euscript}
\usepackage{amsfonts}
\usepackage{amsmath}
\usepackage{color}
\usepackage{subfigure}
\usepackage{epsfig}
\usepackage{subfigure, rotating, bm, array}
\usepackage[pagebackref=false, colorlinks=true]{hyperref}
\hypersetup{
linkcolor=blue,     % color of internal links
citecolor=blue,     % color of links to bibliography
urlcolor=blue}      % color of the url
\begin{document}

\title{Particle motion and tidal force in a non-vacuum-charged naked singularity}

\author{Divyesh P. Viththani}
\email{divyeshviththani@gmail.com}
\affiliation{PDPIAS,
Charotar University of Science and Technology, Anand- 388421 (Guj), India.}

\author{Ashok B. Joshi}
\email{gen.rel.joshi@gmail.com}
\affiliation{PDPIAS,
Charotar University of Science and Technology, Anand- 388421 (Guj), India.}

\author{Tapobroto Bhanja}
\email{tapobroto.bhanja@gmail.com}
\affiliation{PDPIAS,
Charotar University of Science and Technology, Anand- 388421 (Guj), India.}

\author{Pankaj S. Joshi}
\email{psjcosmos@gmail.com}
\affiliation{International Centre for Space and Cosmology, School of Arts and Sciences, Ahmedabad University, Ahmedabad-380009 (Guj), India.}

\date{\today}

\begin{abstract}
We investigate the gravitational field of a charged, non-vacuum, non-rotating, spherically symmetric body of mass $M$ assuming a static solution to the Einstein-Maxwell field equations. We show the characteristics of perihelion precession of orbits in the case of charged naked singularity (CNS) spacetime. Here we discuss some novel features of light-like geodesics in this spacetime. We also discuss the comparative study of tidal force in the null singularity spacetime and charged naked singularity spacetime.
\bigskip

$\boldsymbol{key words}$ : Naked singularity spacetime, Black hole spacetime.
\end{abstract}

\maketitle

%%%%%%%%%%%%%%%%%%%%%%%%%%%%%%%%%%%%%%%%%%%%%%%%%%%%%%%%%%%%%%%
\section{Introduction}
The Event Horizon Telescope (EHT) collaboration, by releasing the first image of the astrophysical black hole at the center of M87 galactic center~\cite{event2019first,akiyama20196first} and the subsequent image of the black hole at the center of our own galaxy - Sagittarius A* (Sgr A*)~\cite{akiyama2022first} has opened up new horizons in the field of astrophysics and black hole physics. Theoretically, there are many possible solutions of Einstein's field equations other than a black hole (BH). A naked singularity (NS) is one such example \cite{PhysRevLett.20.878,perlick}. Null geodesics can escape from such a singularity (i.e., NS) and would be observable to a distant observer under certain conditions of gravitational collapse \cite{Joshi:2011zm, Joshi:2013dva}. The singularity could then be locally or globally visible depending upon the time of formation of trapped surfaces and the apparent horizon. Recent observational studies of modified gravity, suggest alternative models of the compact object\cite{Khodadi:2022pqh, Khodadi:2021gbc, Khodadi:2020gns, KumarWalia:2022ddq}. In the case of NS, most of the studies are focused around the accretion discs \cite{Tahelyani:2022uxw, Patra:2023epx, Kovacs:2010xm}, dark matter \cite{Joshi:2022azj}, shadows cast by them \cite{Kumar:2020ltt, Shaikh:2022ivr, Ghosh:2021txu}, and particle trajectories around them \cite{Bambhaniya:2019pbr, Battista:2022krl}.

If both BH and NS exist in nature, they should have highly distinct features both physically and causally along with diverse astrophysical signatures. Here, we take into consideration two particular spacetimes as examples, namely the future-null singularity (FNS) \cite{Joshi:2023ugm, joshi2020shadow, Paul2020} and charged naked singularity (CNS, which we introduce in this work), in order to explore the possibilities of identifiable observational signatures. The different kinds of physical conditions in the presence of compact objects (BH and NS) would affect the motion of particles and matter around them (resulting in distinct observable signatures), serving as a tool for identifying them.

From an observational standpoint, it might be interesting to note that the UCLA Galactic Centre Group has recently shown that one can effectively use short-period stars (like S2) orbiting around our galaxy's supermassive central compact object to study gravitation theory in the strong field regime \cite{Gillessen:2017jxc}. In \cite{remo2, remo1, Pugliese:2013zma}, it has been shown that for circular motion of 
test particles in Reissner–Nordström (RN) spacetime with vanishing angular momentum shows effects of repulsive gravity. Similar to the RN case we have shown here that the effect of charge could play a significant role in the formation of stable circular orbits even with zero angular momentum.

Further, it is generally believed that a test body approaching a static, spherically symmetric compact object experiences compression in the angular and spaghettification in the radial directions, as in the case of Schwazchild BH and FNS \cite{Madan:2022spd}. In the case of electrically charged compact objects (e.g., RN BH) it has been shown by Crispino \textit{et al.}, \cite{Crispino:2016pnv, Arora:2023ltv } that radial and angular components of the tidal effects experienced by a freely falling test body changes sign at the null hypersurface. It is interesting to note that the charge-to-mass ratio of the charged compact object and the position of the test body determine whether it would experience stretching or compression in a radial or angular direction in the vicinity of a charged compact object. The above facts motivated us to make a comparative study on the effects of tidal forces experienced by a test body in the vicinity of FNS and RN spacetimes in our present work. 

In this article, the Einstein-Maxwell field equations for a static spherically symmetric spacetime with a non-zero energy-momentum tensor (non-vacuum case) are discussed. We analyze three geometrical properties of spacetime: (i) the nature of the precession of charge-less test particles, (ii) the shadow and gravitational lensing property of lightlike geodesics, and (iii) tidal force(s) in CNS spacetime. It follows that it is worthwhile to investigate the theoretical predictions for plausible BH and NS spacetime signatures, as well as their differences and similarities. Understanding a particle's nature of precession in such spacetimes could be a difficult and fruitful problem in general relativity (GR), which is typically handled in terms of the timelike geodesics along which a test particle moves in a particular spacetime. 

Our present work is organized as follows. In Sec.~\ref{spacetime},  we introduce a charge in the FNS spacetime. We study the nature of the spacetime singularity and the energy conditions are explored for the same. In Sec.~\ref{orbit}, we investigate the timelike geodesics and the nature of the precession of a test particle in the CNS spacetime. Our Sec.~\ref{tidal} is dedicated to study the tidal forces (radial and angular) acting on free-falling test bodies in CNS spacetime. In Sec.~\ref{results}, we briefly discuss and summarize the results of our present investigation. Throughout the paper, the signature (-,+,+,+) and the geometric units in which $G = 1 = c$, are used.

\section{Charged Naked Singularity Spacetime (CNS)}\label{spacetime}

\begin{figure}[ht!]
\centering
\includegraphics[scale=.5]{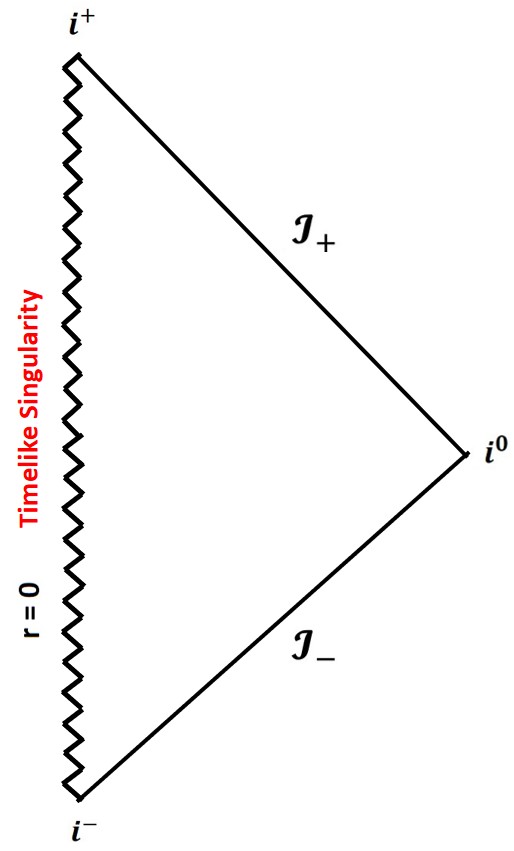}
\caption{Conformal diagram of charged naked singularity spacetime}
\label{PenroseDiagram}
\end{figure}

The metric of a static and spherically symmetric null singularity spacetime can be written as,
\begin{equation}
    ds^2_{FNS} = -\left(1+{\frac{M}{r}}\right)^{-2}dt^2 + \left(1+{\frac{M}{r}}\right)^{2} dr^2
    + r^2d\Omega^2\, ,
    \label{nullmetric}
\end{equation}
where $d\Omega^2= d\theta^2 + sin^2\theta d\phi^2$ and $M$ is the ADM mass of the spacetime. It is shown in ~\cite{joshi2020shadow} that there is a strong null singularity at the center of the spacetime in Eq.~(\ref{nullmetric}).
The Penrose diagram of spacetime can be used to determine the type of spacetime singularity. The Penrose diagram allows us to explain the entire spacetime manifold in a limited-size causal diagram by transforming the temporal and radial coordinates. The compactification of coordinates is as follows:
\begin{eqnarray}
    T = \tan^{-1} (t + r^{*}) + \tan^{-1} (t - r^{*})\nonumber\\
    R = \tan^{-1} (t + r^{*}) - \tan^{-1} (t - r^{*})
    \label{compactcoor}
\end{eqnarray}
where for Eq.~(\ref{nullmetric}),
\begin{equation}
    r^{*} = r -\frac{M^{2}}{r} + 2M\log r . \label{rstar}
\end{equation}
In order to determine the nature of the singularity at $r \to 0$, we need to check the value of $T$ and $R$ as $r \to 0$. From the above Eq.~(\ref{rstar}), it can be verified that as  
$r \to 0$ for any finite value of $t$, $r^{*}  \to - \infty$, clearly suggesting that the singularity thus formed at $r = 0 $ corresponding to the line element defined in Eq.~(\ref{nullmetric}) is null in nature. (Similarly one would get a timelike singularity if for any finite $t$, $r^{*} \to 0$ when $r \to 0$) (See e.g., \cite{dDey}).
There is no event horizon around the singularity. Moreover, this future null singularity (FNS) spacetime satisfies both weak and strong energy conditions ~\cite{joshi2020shadow}. For our present investigation, we introduce an electric  charge in FNS spacetime, and the corresponding action for the electromagnetic field can be expressed as:
\begin{equation}
    S = \int \sqrt{-g}  \left(\frac{R}{16\pi G} - \frac{1}{4}(F_{\mu\nu}F^{\mu\nu})\right) d^{4}x,
\end{equation}
where, $F_{\mu\nu}$ is the usual electromagnetic field tensor. The associated Einstein-Maxwell field equation is given as follows:
\begin{equation}
    R_{\alpha \beta} - \frac{1}{2}R g_{\alpha \beta} = G_{\alpha \beta} = 8\pi(T_{\alpha \beta}^{M} + T_{\alpha \beta}^{EM})\label{EFE},
\end{equation}
where, $T_{\alpha \beta}^{M}$ is the energy-momentum tensor of distributed matter in FNS spacetime given in \cite{Joshi:2023ugm, joshi2020shadow}, while $T_{\alpha \beta}^{EM}$ is the energy-momentum tensor of electromagnetic energy,
\begin{equation}
    T_{\alpha \beta}^{EM} = \frac{1}{\mu_{0}} \left(\frac{1}{4} g_{\alpha \beta}F_{\mu \nu}F^{\mu \nu} - g_{\beta \nu}F_{\alpha \mu}F^{\nu \mu}\right).
\end{equation}
The total energy-momentum tensor of the matter field and charge distribution is given as follows:
\begin{eqnarray}
&& -\, T^0_0 = \frac{M^2 (M+3 r)}{r^2 (M+r)^3}
+ \frac{Q^2}{r^4}, \label{den}\\
&& T^1_1 = - \frac{M^2 (M+3 r)}{r^2(M+r)^3}
- \frac{Q^2}{r^4}, \label{pre}\\
&& T^2_2 =  \frac{3 M^2}
{(M+r)^4}+\frac{Q^2}{r^4}, \label{Tpre}\\ 
&& T^3_3 = \frac{3 M^2}{(M+r)^4}+\frac{Q^2}
{r^4}.\label{mome}
\end{eqnarray}
The most general static and spherically symmetric spacetime metric  is given by: 
\begin{equation}
    ds^2 = - f(r)dt^2 + \frac{dr^2}{g(r)} + r^2(d\theta^2 + \sin^2\theta d\phi^2)\,\,\label{geomerty}.
\end{equation}
By solving the Einstein field equation that is Eqs.~(\ref{EFE}) using the geometry given in Eqs.~(\ref{geomerty}) and total energy-momentum tensor given in Eqs.~(\ref{den}), (\ref{pre}) and Eqs.~(\ref{Tpre}). We get $f(r) = g(r)$, where, 
\begin{equation}
   f(r) = \left(\left(1 +\frac{M}{r}\right)^{-2} + \frac{Q^{2}}{r^{2}}\right).
\end{equation}
This corresponds to a static spherically symmetric body where the charged matter is distributed over the spacetime. The static charge that surrounds such an object contributes to the energy distribution of the spacetime, modifying the spacetime structure. For example, when an electric charge is introduced in a Schwarzschild BH, resulting in the RN BH, the nature of singularity changes, i.e., from spacelike to timelike. However, there is a large possibility that the charge near the singularity plays a significant role in the instability of spacetime singularity. In the CNS case, as shown in Fig.~(\ref{PenroseDiagram}), spacetime singularity is timelike. Here we do not claim that this singularity is stable instead we show that charge in FNS spacetime changes a causal structure similar to the RN case. The corresponding Penrose diagram for the given metric (\ref{geomerty}) is in the Fig.~(\ref{PenroseDiagram}). 

Presented metric (\ref{geomerty}) is asymptotically flat and the inclusion of electric charge in this spacetime is attributed to the form of the chosen action, similar to that of RN black hole spacetime. The corresponding Kretschmann scalar ($K = R_{\alpha \beta \gamma \delta}\, R^{\alpha \beta \gamma \delta}$) and Ricci scalar for this new Charged Naked Singularity (CNS) are: 

\begin{eqnarray}
K &=& \frac{4}{r^8} \Big[-\frac{2\, M Q^2 r^2 \Big(M^3 + 4\,
M^2\, r\, +\,6\, M\, r^2\,+\,12\, r^3\,\Big)}{(M+r)^4}\nonumber\\
&+&\frac{M^2\,r^4}{(M+r)^8} \Big(M^6 + 8\, M^5 r + 26\, M^4 r^2\,
+ 44\, M^3 r^3\nonumber\\
&+&46\, M^2\, r^4\,+\,24\, M\, r^5\,+\,12\, r^6\,\Big)\,+\,14\,
Q^4)\Big],
\end{eqnarray}
and, 
\begin{equation}
   R =  \frac{2 M^3 (M+4 r)}{r^2 (M+r)^4}. 
\end{equation}
We observe that the values of the Kretschmann scalar
and Ricci scalar blow up as $r \rightarrow 0$, 
suggesting the existence of a strong curvature singularity at 
$r = 0$. We do not have any null surface covering the singularity, 
i.e., in simpler terms, there is no event horizon around the 
singularity and hence the spacetime singularity is visible to an 
asymptotic observer.
 
From the above set of equations ((\ref{den})-(\ref{mome})), it can be easily verified that null, weak and strong energy conditions are satisfied:
\begin{equation}
    \rho > 0, \hspace{0.3cm} \rho + p_{r} = 0,\hspace{0.3cm}
\rho + p_{r} + p_{\theta} + p_{\phi} >0. 
\end{equation}
The CNS spacetime also satisfies the dominant energy condition, viz., 
$p_r + p_\theta > 0$. 
%Further we observe that $\rho = -\, p_r$ and the spacetime is seeded by an anisotropic fluid with the anisotropy in the pressure being
%\begin{equation}
%p_r - p_\theta =  -\frac{M^3 (M+4 r)}{r^2 (M+r)^4}-\frac{2 Q^2}{r^4},
%\end{equation}
%which blows up as $r \to 0$ and the corresponding equation of state
%($\omega$)is given by
%\begin{eqnarray}
 %  \omega &=& \frac{p_r + p_\theta + p_\phi}{3\,\rho}\nonumber\\
%&=& \left(\frac{(M+r)^3}{3 \big[M^2 (M+3 r) \left(Q^2+r^2\right) 
%+ Q^2 r^2 (M+r)\big]}\right)\nonumber\\
%&\times&\left(Q^2-\frac{M^2 r^2 \left(M^2+4 M r-3 r^2\right)}
%{(M+r)^4}\right)
%\end{eqnarray}
%such that $\omega \to 1/3$ as $r \to 0$.
%Interestingly, as the pressure is anisotropic, we can safely 
%assume that this spacetime geometry is not seeded by any minimally coupled scalar fields (or vector fields).

%%%%%%%%%%%%%%%%%%%%%%%%%%%%%%%%%%%%%%%%%%%%%%%%%%%%%%%%%%%%%%%%%%%%%%%%%%%%%%%%%%%%%%%%%%%%%%%%%%%%%%%%%%%%%%%%%%%%%%
\begin{figure*}[ht!]
\centering
\subfigure[$M = 1$, $L = 4$, $\mathcal{E} = 0.99$ and $Q = 2.5$]
{\includegraphics[width=45mm]{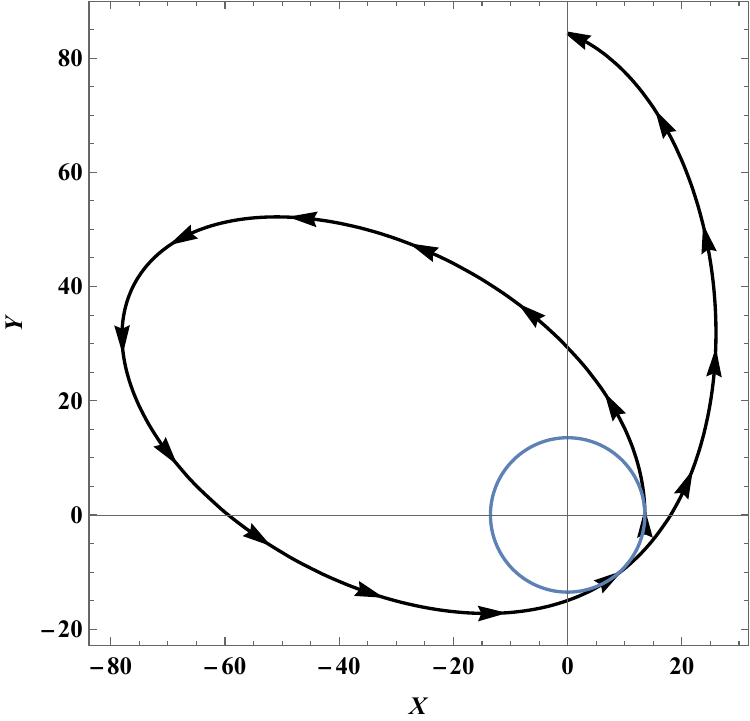}\label{negp1}}
\hspace{0.5cm}
\subfigure[$M = 1$, $L = 4$, $\mathcal{E} = 0.99$ and $Q = 0.4$]
{\includegraphics[width=52mm]{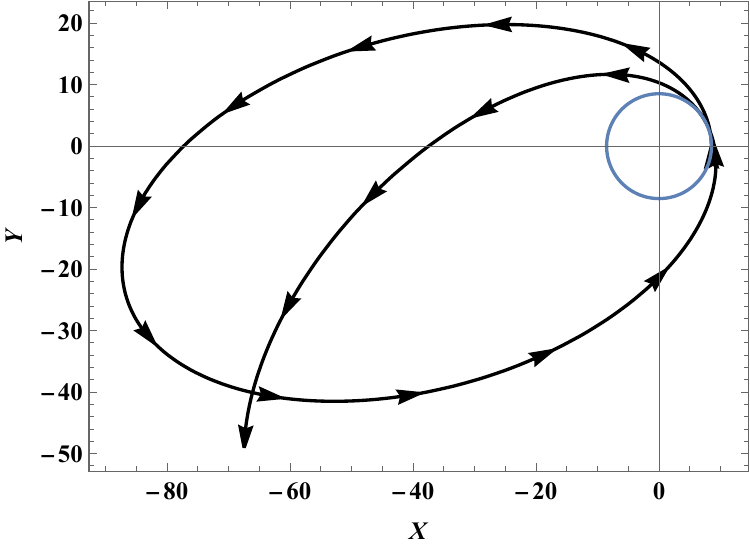}\label{posp1}}
\hspace{0.5cm}
\subfigure[$M = 1$, $L = 4$, $\mathcal{E} = 1.01$ and $Q = 0.4$]
{\includegraphics[width=42mm]{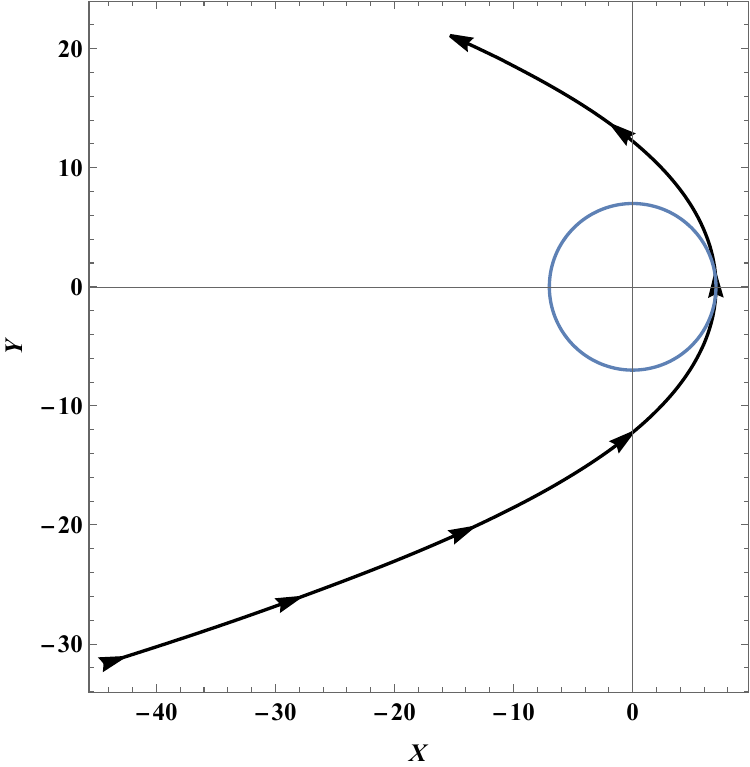}\label{refl2}}
 \caption{ In this figure, particle orbits in charged null singularity spacetime are shown.  It can be seen that for $M = 1$, $L = 4$, $\mathcal{E} = 0.99$, and $Q = 2.5$, the angular distance traveled by the particle to reach one perihelion point to another perihelion point is less than $2\pi$, whereas, for $Q = 0.4$, it reaches after $2\pi$ rotation. Unbound orbits are shown as (\ref{refl2}).}\label{prece}
\end{figure*}
%%%%%%%%%%%%%%%%%%%%%%%%%%%%%%%%%%%%%%%%%%%%%%%%%%%%%%%%%%%%%%%%%%%%%%%%%%%%%%%%%%%%%%%%%%%%%%%%%%%%%%%%%%%%%%%%%%%%%%

\section{Timelike and Lightlike geodesics in Charged naked singularity spacetimes }\label{orbit}

As discussed in  Sec. \ref{spacetime}, CNS spacetime is asymptotically flat and harbors a central strong curvature singularity which is not covered by an event horizon. It has temporal and rotational (about the azimuthal angle) symmetries leading  to two associated Killing vectors: $\xi^{\mu}_t = (1, 0, 0, 0)$ and  $\xi^{\mu}_\phi = (0, 0, 0, 1)$, such that $\xi^{\mu}_t \, \partial_\mu = \partial_t$ and  $\quad \xi^{\mu}_{\phi} \, \partial_\mu = \partial_\phi$. The corresponding conserved quantities: 
\begin{equation}
    \mathcal{E} = \dot t \, f (r), \quad
     l = \dot \phi \, r^2, 
\end{equation}
where $\mathcal{E}$ and $l$ are the energy per unit rest mass and conserved angular momentum for a freely falling particle in CNS spacetime, respectively. Here, the `overdot' means a derivative with respect to the `proper time' ($\tau$) of the particle. Using the normalization condition (for timelike particles), $u_\alpha \, u^\alpha = - 1$, the total energy ($E$) of the freely falling massive particle can be written as,
\begin{equation}
   E = \left(\frac{dr}{d\tau}\right)^2 + V_{eff}(r),\,\, 
    \label{totalE}
\end{equation}
where $E = \mathcal{E}^{2}$. The corresponding  effective potential in which a test particle of mass $m$ moves in the equatorial plane for CNS is given by: 
\begin{equation}
     (V_{eff})_{CNS}= f(r)\left(1 + \frac{l^2}{r^2}\right). 
    \label{veffsch}
\end{equation}
For a bound orbit, at maximum radial distance ($r_{max}$) and minimum radial distance ($r_{min}$), these condition must be satisfied: 
 \begin{eqnarray}
   V_{eff}(r_{min})=V_{eff}(r_{max})=E\, , \,\, \nonumber\\
   E-V_{eff}(r)>0\, ,\,\,\,\forall \,\, r\in (r_{min},r_{max}).
   \label{bound}
\end{eqnarray}

For a stable circular orbit of a test particle around a supermassive object, at $r_{min} = r_{max} = r_{c}$, the energy of a particle is given by $V_{eff}(r_c) = E$ and $V_{eff}^{\prime}(r_c)=0$ and $V_{eff}^{\prime\prime}(r_c)>0$, where $r_c$ is the radius of the stable circular orbit, and `prime' ($\prime$) represents a derivative with respect to the radial coordinate $r$. Using these facts, the  innermost stable circular orbit (ISCO) in CNS spacetime for $l=0$ is at,
\begin{widetext}
    \begin{equation}
        r_{c} = r_{ISCO} = \frac{Q^2}{4M} + \frac{1}{2} \sqrt{Z-X+Y} + \frac{1}{2} \sqrt{2 Z+X - Y + \frac{24M Q^{2} + \frac{12Q^{4}}{M} + \frac{Q^{6}}{M^{2}}}{4\sqrt{Z - X + Y}}}\label{risco},
    \end{equation}
    where,
    \begin{eqnarray}
        X &=& \frac{4 (2/3)^{1/3} M^3 Q^2}{\left(9 M^5 Q^4 + \sqrt{3} \sqrt{256 M^{12} Q^6 + 27 M^{10} Q^{8}}\right)^{1/3}}, \nonumber\\
        Y &=& \frac{\left(9 M^5 Q^4 + \sqrt{3} \sqrt{256 M^{12} Q^6 + 27 M^{10} Q^{8}}\right)^{1/3}}{(3)^{2/3} (2)^{1/3} M}, \nonumber\\
        Z &=& 2 Q^2 + \frac{Q^4}{(4 M^2)}.
    \end{eqnarray}
\end{widetext}
From Eq.~(\ref{risco}), if $Q=0$, $r_{ISCO}=0$, i.e., in FNS spacetime ISCO forms at $r =0$. From the expression of effective potential in (\ref{veffsch}), we can derive the expression of $l$ and $E$ for circular timelike geodesics,
\begin{eqnarray}
   E = \frac{\left(\frac{Q^2}{r^2} + \frac{r^2}{(M + r)^2}\right)}{\sqrt{\frac{2 Q^2}{r^2} + \frac{r^3}{(M + r)^3}}},\label{lecns}\quad
   l = \frac{r^{2} \sqrt{-\frac{Q^2}{r^4} + \frac{M}{(M + r)^3}}}{\sqrt{\frac{2 Q^2}{r^2} + \frac{r^3}{(M + r)^3}}} .\label{hecns}
\end{eqnarray}
Using the expressions of conserved quantities for circular geodesic, one can show that no circular orbit is possible in the range: $0\leq r <  r_{ISCO}$. However, for an innermost stable circular orbit, need to satisfy $ (V_{eff}^{\prime\prime})_{CNS}>0$ along with the above two conditions in Eq.~(\ref{lecns}). In terms of the ADM mass $M$ and charge $Q$, with condition $l=0$, the expression of $ (V_{eff}^{\prime\prime})_{CNS}$ becomes:
\begin{eqnarray}
   (V_{eff}^{\prime\prime})_{CNS}&=&\frac{6 Q^2}{r^4} + \frac{2 M (M - 2 r)}{(M + r)^4}\,\, .
   \label{isco}
\end{eqnarray}
For ISCO at $r = r_{ISCO}$ and $Q>0$, Eq.~(\ref{isco}) gives a positive finite value which suggest $r_{ISCO}$ is stable. The minimum value of energy for a stable circular orbit is given by:
\begin{equation}
    E_{ISCO} = \sqrt{\frac{(r_{ISCO})^2 (2 M + r_{ISCO})}{(M + r_{ISCO})^3}}.
\end{equation}
The shape of the orbit of a test particle in the CNS spacetime can be derived from Eq.~(\ref{totalE}), 
\begin{equation}
    \frac{d\phi}{dr}= \frac{l}{r^2 \sqrt{2(E-V_{eff}(r))}}\,\, ,
    \label{orbitgen}
\end{equation}
where $\phi$ and $r$ are the azimuthal and radial coordinates, respectively. Now, using the above equation one can define the following second-order differential equations for timelike geodesic in CNS spacetime, 
 \begin{eqnarray}
   \frac{d^2u}{d\phi^2} &=& \frac{M}{l^{2}
   (1+M u)^{3}} - u \left(\frac{Q^{2}}{l^{2}} + 
   \frac{1}{(1 + M u)^{2}}\right)\nonumber\\
   &+& \frac{M u^{2}}{(1+M u)^{3}}
   - 2 Q^{2} u^{3},
   \label{orbiteqcns}
\end{eqnarray}
where $u=\frac1r$. From the orbit Eq. (\ref{orbiteqcns}), one can get information about the shape of orbits in charged naked singularity spacetime and can compare them to RN-BH spacetime.

%%%%%%%%%%%%%%%%%%%%%%%%%%%%%%%%%%%%%%%%%%%%%%%%%%%%%%%%%%%%%%%%%

\subsection{Approximate solution of orbit 
equations in Charged naked singularity spacetime}\label{approximation}

The orbit Eq in ~(\ref{orbiteqcns}) is highly complicated and is difficult to analyze analytically. Hence to check the nature of the precession of orbit(s) we Taylor expand Eq. ~(\ref{orbiteqcns}). The corresponding orbit equation for a test particle that is revolving in the weak gravity region now becomes: 
\begin{multline}
\frac{d^2u}{d\phi^2} = \frac{M}{l^{2}} - 
\left(1 + \frac{3 M^2}{l^2} + \frac{Q^2}{l^2}
\right) u + \left(3 M + \frac{6 M^3}{l^2} 
\right) u^2 +\\ - \left(6 M^2 + \frac{10 M^4}
{l^2} + 2 Q^2\right) u^3 + \mathcal{O}(u^4). \label{apporbit}
\end{multline}
With this approximation, one can get important information about the nature and shape of bound orbits which are difficult to get from the original orbit equation  ~(\ref{orbiteqcns}). Further, we consider two more approximations: (i) an approximation on the radial distance which is a weak gravity approximation, and (ii) by considering the small values of eccentricity ($e$). This approximate solution of the orbit equation method is extensively discussed in \cite{Bambhaniya:2019pbr}. We can write the first-order eccentricity approximate solution \cite{Struck:2005hi, Struck:2005oi} for the given orbit Eq.~(\ref{apporbit}) as follows:
\begin{equation}
\tilde{u}=\frac{1}
{p}\left[1+e\cos(m_0\phi)+O(e^2)\right]\,\, ,
\label{orbitsch1}
\end{equation}
where $m_0$ and $p$ are positive real values and $\tilde{u}=M u$. If $m_0 < 1$, it implies that for one full periodic rotation $\phi>2\pi$. When the precession angle $\delta =\phi - 2\pi$ gives a positive value, the nature of the precession of the test particle trajectories is called a positive precession.  Schwarzschild spacetime always shows positive precession for all parameters of space (See, e.g., \cite{Bambhaniya:2019pbr}). Similarly $m_0 > 1$ gives the negative value of $\delta$ and is known as negative precession, while for $\delta = 0$ when $m_0 = 1$ is the Newtonian case. 

Using the $3^{rd}$ order approximated solution of $u$ and considering the eccentricity approximation as given in Eq.~(\ref{orbitsch1}), we can get the following expression of $p$ and $m$ for CNS spacetime:
\begin{eqnarray}
p &=& \frac{ \beta }{3 \alpha } - \frac{2^{1/3} \psi}
{3 \alpha \left(\Sigma + \sqrt{4\psi^{3} + 
\Sigma^{2}}\right)^{1/3}}\nonumber\\
&+& \frac{2^{2/3} \left(\Sigma + 
\sqrt{4\psi^{3} +
\Sigma^{2}}\right)^{1/3}}{6 \alpha },\nonumber\\
%\begin{equation}
m_0 &=&  \sqrt{\beta - \frac{2\gamma}{p} + \frac{3 \omega}
{p^{2}}},\label{mcns}
\end{eqnarray}
%\end{equation}
Where, $\psi = (-\beta^{2} + 3 \alpha \gamma),\,\,\ \Sigma 
= (2 \beta ^{3} - 9\alpha \beta \gamma + 27\alpha^{2} \omega)$, and 
\begin{eqnarray}
&& \gamma = \left(3 M + \frac{6 M^3}{l^2} \right), \quad \,\,
\beta = \left(1 + \frac{3M^{2}}{l^{2}} + 
\frac{Q^{2}}{l^{2}}\right),\nonumber\\
&& \alpha = \frac{M}{l^{2}},
\hspace{.9cm} \omega = \left(6 M^2 + \frac{10 M^4}
{l^2} + 2 Q^2\right).
\end{eqnarray}

The above set of equations determine the nature of precession for two the spacetimes: FNS ($Q = 0$) and CNS ($Q > 0$): 
\begin{enumerate}
    \item {\bf{For $Q=0$:}} the value of $m_0$ given in Eq.~(\ref{mcns}) is always smaller the 1 suggesting that FNS spacetime always possesses positive precession like Schwarzschild spacetime.
    \item {\bf{For $Q>0$:}} From  Eq.~(\ref{mcns}), when the charge to mass ratio is greater than 1.3, (i.e., $\frac{Q}{M} > 1.3$) we have $m_0>1$ and thus negative precession while for  $0< \frac{Q}{M} <1.3$ we have $m_0 < 1$ and thus positive precession of the test particle around the supermassive compact object.
\end{enumerate}
The above-given value of charge-to-mass ratio $1.3$ is an approximate value that can be fine-tuned by taking higher order correction.

%%%%%%%%%%%%%%%%%%%%%%%%%%%%%%%%%%%%%%%%%%%%%%%%%%%%%%%%%%%%%%%%%%%
%%%%%%%%%%%%%%%%%%%%%%%%%%%%%%%%%%%%%%%%%%%%%%%%%%%%%%%%%%%%%%%%%%%
%%%%%%%%%%%%%%%%%%%%%%%%%%%%%%%%%%%%%%%%%%%%%%%%%%%%%%%%%%%%%%%%%%%%%%%%%%%%%%%%%%%%%%%%%%%%%%%%%%%%%%%%%%%%%%%%%%%%%%%%%%%%%%%%%%%

\begin{figure*}[ht!]
\centering
\subfigure[]
{\includegraphics[width=62mm]{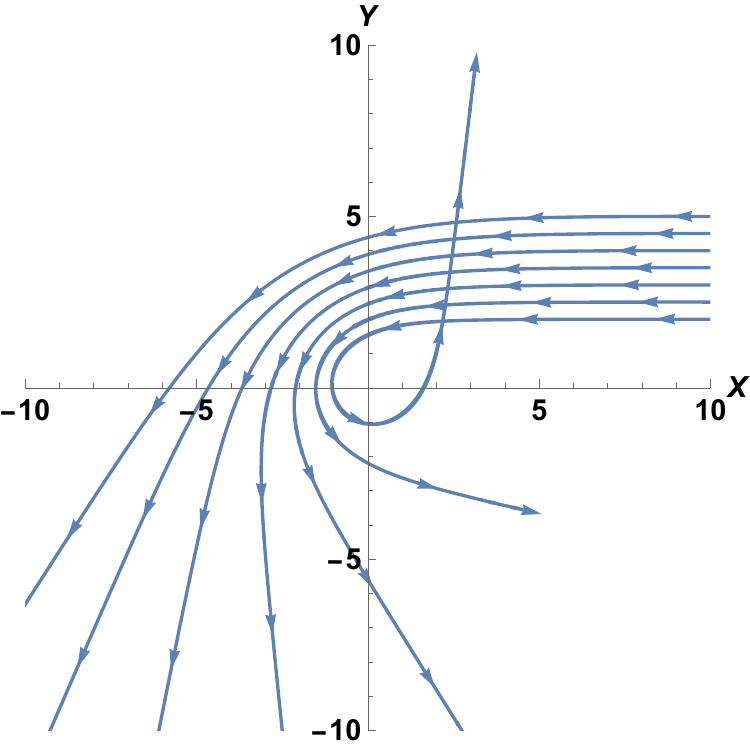}\label{negp2}}
\hspace{1cm}
\subfigure[]
{\includegraphics[width=62mm]{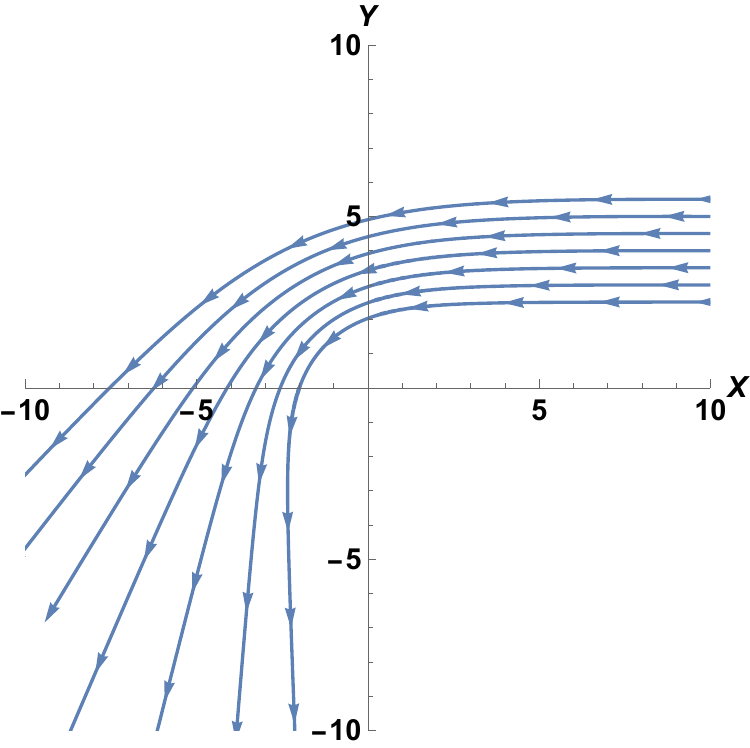}\label{negp3}}
\hspace{1cm}
\subfigure[]
{\includegraphics[width=62mm]{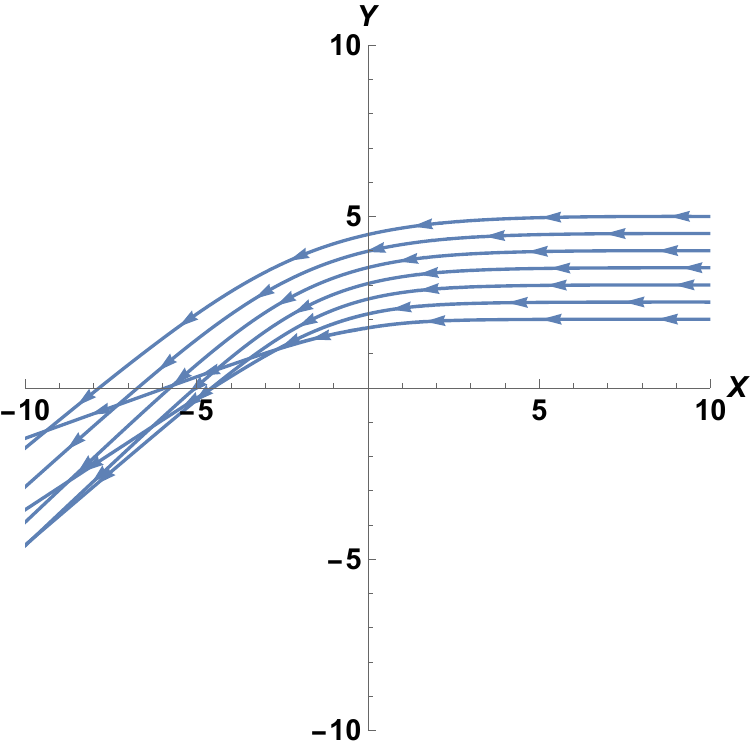}\label{posp2}}
\hspace{1cm}
\subfigure[]
{\includegraphics[width=62mm]{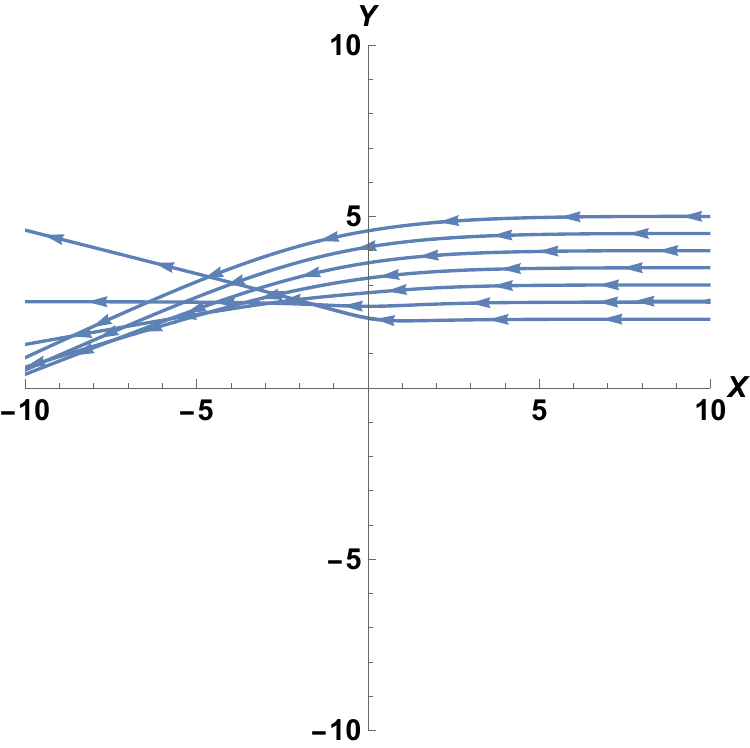}\label{posp3}}
 \caption{Gravitational Lensing when: 
 (a) $Q<<M , (Q=0.01)$, (b) $Q<M , (Q=0.5)$,  (c) $Q=M , (Q=1)$, (d)  $Q>M , (Q=1.5)$}\label{lensing}
\end{figure*}

%%%%%%%%%%%%%%%%%%%%%%%%%%%%%%%%%%%%%%%%%%%%%%%%%%%%%%%%%%%%%%%%%%%%%%%%%%%%%%%%%%%%%%%%%%%%%%%%%%%%%%%%%%%%%%%%%%%%%%%%%%%%%%%%%%%%%%%%%%%%%%%%%%%%%%%%%%%%%%%%%

\subsection{Lightlike geodesics}

As discussed earlier, CNS is spherically 
symmetric and static, hence for the $\theta = \pi/2$ (i.e., for the
equatorial plane) we have  (with $f(r) = g(r)$) \cite{joshi2020shadow},
\begin{equation}
    \frac{1}{b^2}=\frac{1}{l^2}\left(\frac{dr}{d\lambda}\right)^2+W_{eff},\label{weff}
\end{equation}
where $W_{eff}=f(r)/r^2$ and $b$ is impact parameter defined by
$b = {\mathcal{E}}/{l}$. To obtain Eq.~(\ref{weff}), we use the condition 
$k_\mu k^\mu=0$, where $k_\mu$ is the null four-velocity. The stable 
and unstable orbits of photons can be investigated based on the 
nature of the effective potential of the spacetime. For an unstable 
circular orbit the effective potential should have an extrema at 
a finite radius, $r = r_{ph}$. The sphere that corresponds to this particular radius, $r_{ph}$ is known as the photon sphere. One can obtain the radius of 
the photon sphere by the following two conditions:
\begin{equation}
\frac{dW(r_{ph})}{dr} = 0, \qquad \quad
\frac{d^2W(r_{ph})}{dr^2} < 0.
\end{equation}
We would have a photon sphere at $r_{ph}$ if the above two conditions 
are satisfied for any spacetime. One can find the turning points of 
null geodesics by:
\begin{equation}
W_{eff}=\frac{\mathcal{E}^2}{l^2}=\frac{1}{b^2_{tp}}.\label{Weff}
\end{equation}
where,
\begin{equation}
b_{tp}=\frac{r_{tp}}{\sqrt{f(r_{tp})}}, 
\end{equation}
and ${r_{tp}}$ is the radius of the turning points. If there is only a single extremum value of the effective potential of null geodesics for any spacetime and simultaneously, if the extremum value corresponds to the maximum value of the effective potential, then the minimum value of the impact parameter of the turning point becomes equal to the impact parameter of the photon sphere, i.e.,  $b_{tp} = b_{ph}$, ($b_{ph} = $ impact parameter of the corresponding photon sphere). As a result, incoming light-like geodesics from a source at infinity, will not reach an asymptotic observer for an impact parameter $b < b_{ph}$. These geodesics will be trapped inside the photon sphere and hence, the null geodesics coming from behind the compact object would create a  shadow of radius $b_{ph}$ for the asymptotic observer (due to the presence of the photon sphere). On the contrary, if there is no photon sphere for any spacetime and the effective potential is diverging at the origin, no shadow would be formed (for that particular spacetime). This is the case with 
CNS, (we \textit{do not} get any shadow because of such geometric structure).
However, as can be seen in Fig.~(\ref{lensing}), lensing of CNS spacetime suggests that if such objects exist in nature, they would be highly luminous objects.

%%%%%%%%%%%%%%%%%%%%%%%%%%%%%%%%%%%%%%%%%%%%%%%%%%%%%%%%%%%%%%%%%%%%%%%%%%%%%%%%%%%%%%%%%%%%%%%%%%%%%%%%%%%%%%%%%%%%%%%%%%%%%%%%%%%%%%%%%%%%%%%%%%%%%%%%%%%%%

\section{Tidal Force}\label{tidal}
In this section, we investigate the effect of tidal force on a
test body near CNS and compare them with Schwarzchild (SCH), RN, and FNS spacetimes, respectively. To investigate tidal force in the framework of the general theory of relativity we analyze the equation of geodesic deviation:

\begin{equation}
    \frac{D^2\eta^\mu}{D\tau^2}-R^\mu_{\nu \rho \sigma}v^\nu v^\rho 
    \eta^\sigma =0,\label{radialtidal}
\end{equation}
where $R^\mu_{\nu \rho \sigma}$ and $v^\nu$ are the Riemann 
curvature tensor and unit tangent vector of the geodesic respectively and $\eta ^\mu$ is the geodesic deviation vector. Any non-zero gradient in the gravitational field implies that each point on the geodesics has a different curvature and thus, at each point of the test body will follow a unique geodesic, leading to stretching and/or squeezing known as tidal effect. To find these stretching and squeezing effects, we use the Jacobi field (a vector field along a geodesic), which is the separation between two infinitesimally close geodesics. Note that an observer's perspective on the relative spatial acceleration of the two particles is useful in understanding the physical implications of the geodesic deviation effect.

In the Jacobi fields, we first define the tetrad components of
a free-falling frame: 
\begin{eqnarray}
&&\hat{e}^\mu_{\hat{0}} = \Biggl\{ \frac{E}{f(r)},-\sqrt{E^2-
f(r)},0,0\Biggl\},\nonumber\\
&&\hat{e}^\mu_{\hat{1}}= 
\Biggl\{ \frac{-\sqrt{E^2-f(r)}}{f(r)},E,0,0\Biggl\},\nonumber\\
&& \hat{e}^\mu_{\hat{2}}=\Biggl\{0,0,\frac{1}{r},0\Biggl\}, \quad
\hat{e}^\mu_{\hat{3}}= \Biggl\{ 0,0,0,\frac{1}{r\sin\theta}\Biggl\}.\label{tetrad}
\end{eqnarray}
%Tidal force can be described by the relation between two infinitesimally close free-falling particles,
In the instantaneous rest frame (IFR), Eq.~(\ref{radialtidal}), can be expressed as
\begin{equation}
\frac{d^2\,\eta^{\hat{\alpha}}}{d\tau^2} = R^{\hat{\alpha}}_{ \hat{0} 
\hat{0} \hat{\gamma}} \, \eta^{\hat{\gamma}},\label{Rhat}
\end{equation}
where, 
$R^{\hat{\alpha}}_{ \hat{\beta} \hat{\gamma} 
\hat{\delta}}=R^a_{bcd}e^{\hat{\alpha}}_a e_{\hat{\beta}}^b 
e_{\hat{\gamma}}^c e_{\hat{\delta}}^d$.
Considering the vectors are parallelly transported along the 
geodesic and exploring the above equations (\ref{tetrad}) and 
(\ref{Rhat}), we obtain the relative acceleration between two nearby particles in radial and tangential directions as follows:
\begin{equation}
\frac{D^2 \eta^{\hat{r}}}{D\tau^2}=-\frac{f^{\prime \prime}}
{2}\eta^{\hat{r}},\qquad
\frac{D^2 \eta^{\hat{i}}}{D\tau^2}=-\frac{f^{\prime}}
{2r}\eta^{\hat{i}},
\end{equation}
where $i = \theta, \phi$. Substituting the metric component 
$ f(r)$ etc., we obtain for radial tidal force, 
\begin{equation}
\frac{D^2 \eta^{\hat{r}}}{D\tau^2}=-\Biggl(\frac{M(M-2r)}
{(M+r)^4}+\frac{3Q^2}{r^4}\Biggl)\eta^{\hat{r}},\label{radial}
\end{equation}
and for the angular part, we have: 
\begin{equation}
\frac{D^2 \eta^{\hat{i}}}{D\tau^2}=\Biggl(\frac{Q^2}{r^4}-\frac{M}
{(M+r)^3}\Biggl)\eta^{\hat{i}}.\label{tangential}
\end{equation}
The above two equations represent the tidal force for a radially 
free-falling frame in the CNS spacetime. Also one can observe from the equation that tidal force in this spacetime metric depends on the mass and charge of the compact object. Interestingly, the radial and angular components of tidal force vanish for some particular radius (depending upon $M$ and $Q$), in contrast to the Schwarzschild spacetime while similar to  RN spacetime.

\subsection{Radial Tidal Force}

\begin{figure}[ht!]
\centering
\includegraphics[scale=.55]{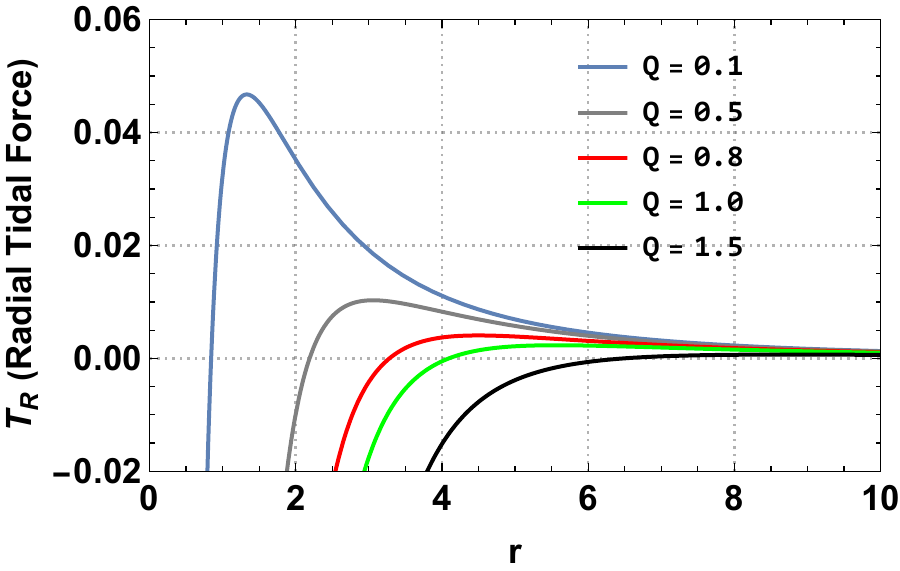}
\caption{Radial Tidal Force ($T_R$) v/s. radial distance $r$}
\label{Radial}
\end{figure}

We observe from Eq.~(\ref{radial}), that the radial tidal force for CNS spacetime goes to negative infinity when the radial distance tends to zero from the spacetime singularity (i.e., $r \to 0$), which is similar to RN spacetime \cite{Crispino:2016pnv}, leading to infinite compression in contrast to the Schwarzchild spacetimes, (where we have infinite spaghettification). The radial component of tidal force vanishes for $r \to \infty$ and as we move towards the singularity, at some  
particular radius, the radial component achieves a local maxima (depending on  specific values of $M$ and $Q$)
after which it falls, goes zero again and proceeds to
negative infinity.
When the radial component vanishes for a finite positive value of r, any local 
observer can observe 
singularity without the  influence of radfial tidal force. As we 
increase the value of the charge $Q$, local maxima shifts right as 
the particle falls radially from infinity.

\begin{figure*}[ht!]
\centering
\subfigure[]
{\includegraphics[width=52mm]{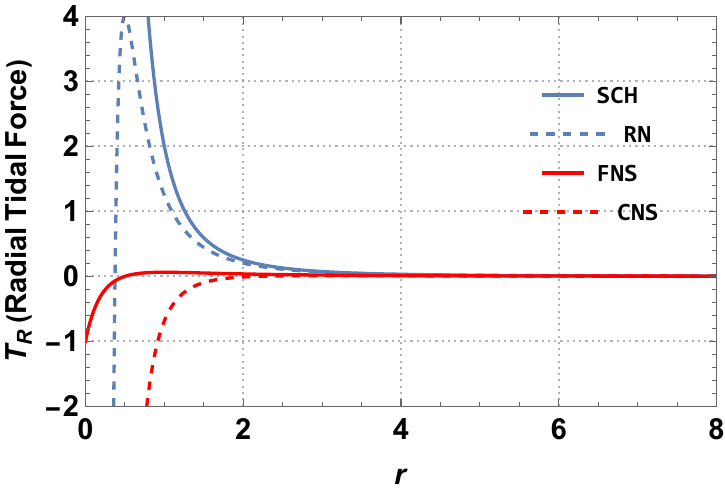}\label{QlessM1}}
\hspace{0.5cm}
\subfigure[]
{\includegraphics[width=52mm]{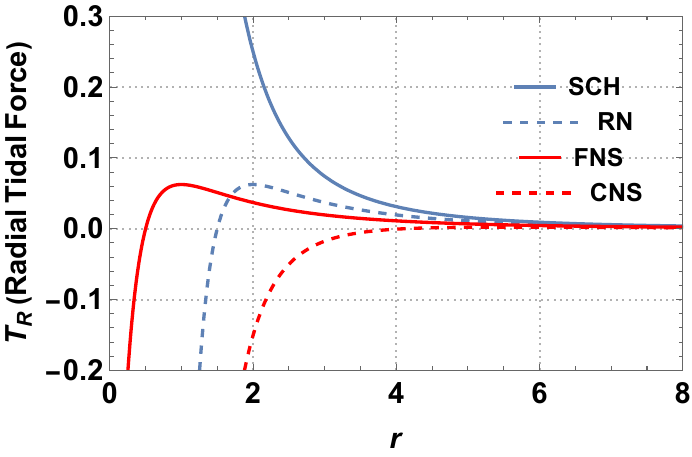}\label{QeqM1}}
\hspace{0.5cm}
\subfigure[]
{\includegraphics[width=52mm]{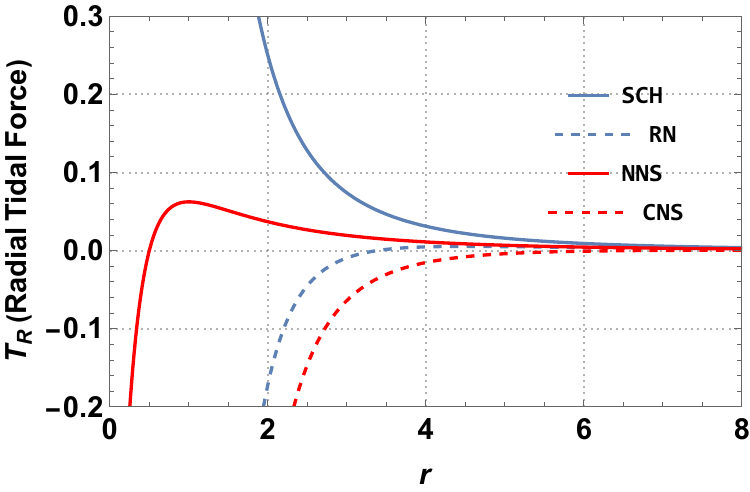}\label{QgraterM1}}\\
\subfigure[]
{\includegraphics[width=52mm]{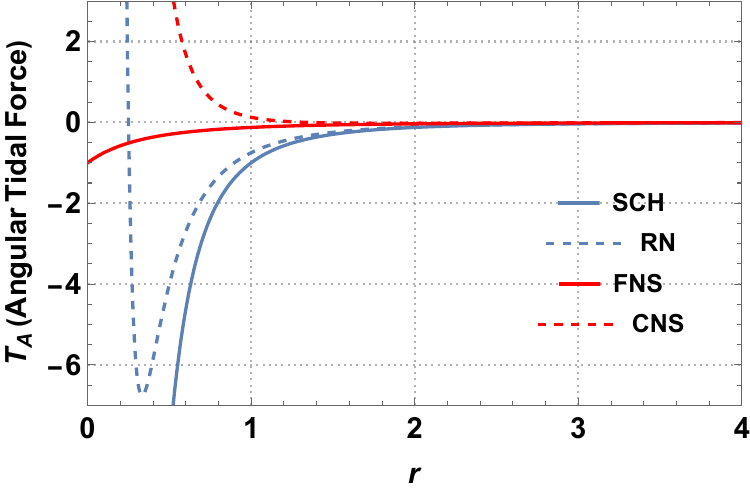}\label{QlessM}}
\hspace{0.5cm}
\subfigure[]
{\includegraphics[width=52mm]{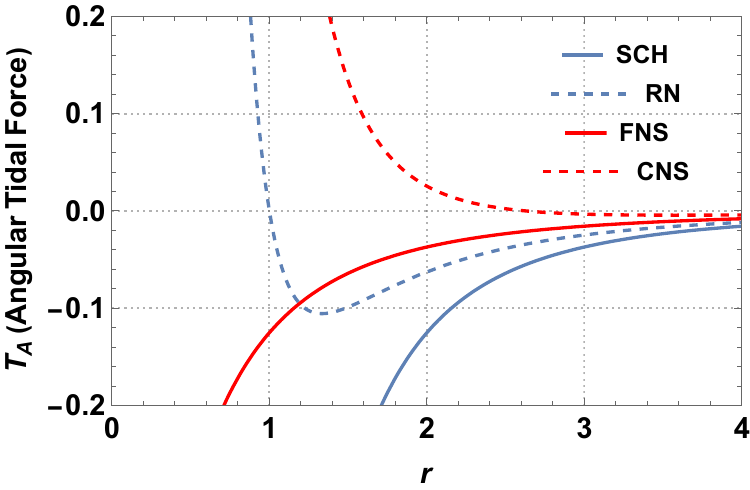}\label{QeqM}}
\hspace{0.5cm}
\subfigure[]
{\includegraphics[width=52mm]{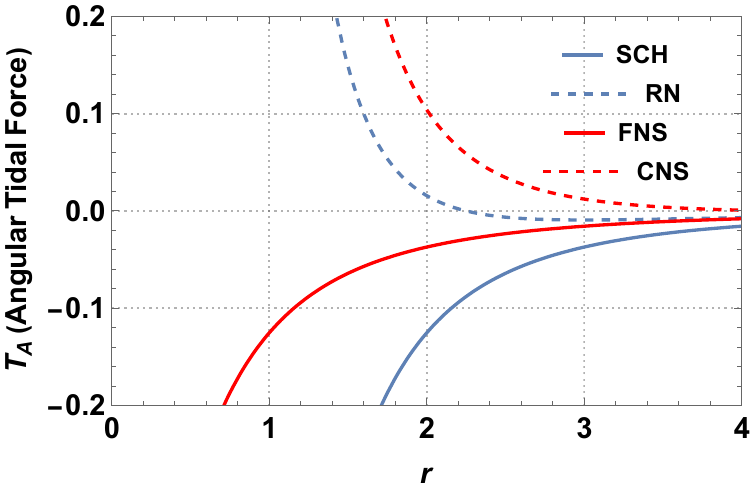}\label{QgraterM}}
 \caption{(a) Radial Tidal force for $Q < M, \, (M = 1, Q = 0.5)$ (b) Radial Tidal force for $Q = M = 1$ (c) Radial Tidal force for $Q > M, \, (M = 1, \, Q = 1.5)$ (d) Angular Tidal force for $Q < M, \, (M = 1, Q = 0.5)$ (e) Angular Tidal force for $Q = M, \, (M = 1 = Q)$ (f) Angular Tidal force for $Q > M, \, (M = 1, Q = 1.5)$}\label{tidalcompare1}
\end{figure*}

%\end{itemize}
\subsection{Angular Tidal Force}

\begin{figure}[ht!]
    \centering
    \includegraphics[scale=.65]{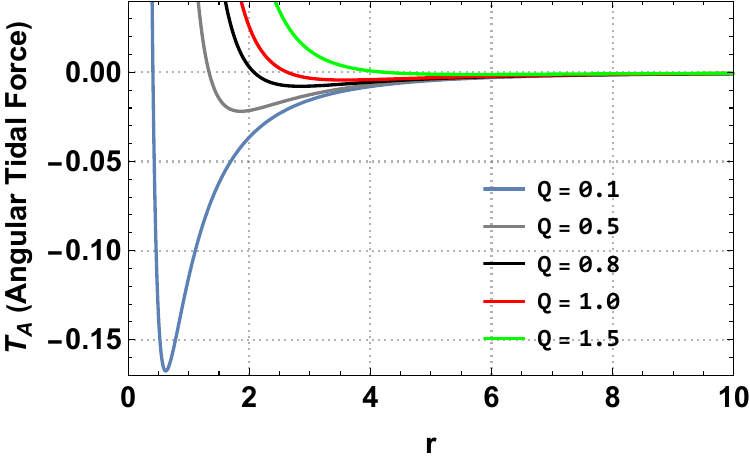}
    \caption{Angular tidal force ($T_A$) v/s. radial distance $r$}
    \label{Angular}
\end{figure}

%%%%%%%%%%%%%%%%%%%%%%%%%%%%%%%%%%%%%%%%%%%%%%%%%%%%%%%%%%%%%%%%%%%%%%%%%%%%%%%%%%%%%%%%%%%%%%%%%%%%%%%%%%%%%%%%%%%%%%

%%%%%%%%%%%%%%%%%%%%%%%%%%%%%%%%%%%%%%%%%%%%%%%%%%%%%%%%%%%%%%%%%%%%%%%%%%%%%%%%%%%%%%%%%%%%%%%%%%%%%%%%%%%%%%%%%%%%%%
%%%%%%%%%%%%%%%%%%%%%%%%%%%%%%%%%%%%%%%%%%%%%%%%%%%%%%%%%%%%%%%%%%%%%%%%%%%%%%%%%%%%%%%%%%%%%%%%%%%%%%%%%%%%%%%%%%%%%%

As can be seen from Fig.~(\ref{Angular}), we observe that similar to the radial component, the angular part of the tidal force vanishes as $r \to \infty$. The 
angular component becomes zero for a certain finite positive value of $r$ and
 we have local minima (depending upon the combination of $Q$ and $M$) beyond which the angular component
goes to positive infinity (as we move towards the spacetime singularity),
signifying spaghettification similar to RN spacetime.

\subsection{Comparison of Tidal force between CNS, RN, FNS, and Schwarzchild spacetimes}

When we compare the nature of the tidal forces in CNS with 
Schwarzchild, RN, and FNS spacetimes, we find some distinct features.
We observe that for the radial component of the tidal forces, there is an 
infinite compression for CNS similar to RN spacetime,  and 
in contrast to Schwarzchild and FNS spacetimes. 
It is to be noted that in CNS, there would be small spaghettification
followed by infinite compression for smaller values of $Q$ 
(i.e., for $Q$ slightly greater than $M$). As we increase the value of
$Q$, spaghettification decreases and  for 
large values of $Q$ there would be no spaghettification (in contrast to 
FNS and Schwarzchild spacetimes). In CNS spacetime, there is a rapid change 
in spaghettification as we increase the value of $Q$ and for large 
values of $Q$, there would be no spaghettification, similar to RN. 
%However, in CNS spacetime, the spaghettification is much smaller
%w.r.t. RN for smaller values of $Q$.
In FNS, there is a finite spaghettification followed by a finite 
compression while there is infinite spaghettification and no 
compression in Schwarzchild spacetime as the test body approaches
the spacetime singularity.
For all the three cases,  $Q > M \quad Q = M$  and $Q < M$, the 
compression sets-in much earlier w.r.t. RN and FNS spacetimes

As for the angular part, we observe that in CNS there is a local 
minima for $Q \le  M$ followed by infinite compression. As we increase $Q$, we observe that the value of local minima decreases and for a large value of $Q$, there is only infinite spaghettification and no compression (similar to  RN  spacetime). In FNS spacetime, we have finite compression while in Schwarzchild we have infinite compression.

%%%%%%%%%%%%%%%%%%%%%%%%%%%%%%%%%%%%%%%%%%%%%%%
%%%%%%%%%%%%%%%%%%%%%%%%%%%%%%%%%%%%%%%%%%%%%%%%

%\newpage
\section{Results and Discussion}\label{results}

In this present work, we introduce and analyze the CNS spacetime which is a solution of Einstein-Maxwell field equations. We have mainly focused on the dynamics of particle trajectories and tidal force in CNS spacetime. The results from this study can be summarised as follows:

\begin{itemize}
    \item In the timelike geodesics, we observe that the nature of the precession of particles changes when the charge-to-mass ratio is larger than $1.3$ ($Q/M > 1.3$), the angular distance traveled by the particle to reach one perihelion point to another perihelion point is less than $2\pi$, while for $Q/M < 1.3$, it is larger than $2\pi$. It is interesting to note that for CNS, in contrast to FNS spacetime, ISCO does not occur at $r = 0$, rather it occurs at some finite positive value of $r$ which is a function of the mass $M$ and charge $Q$ as shown in (\ref{risco}) for $l = 0$. It is interesting to note that as we do not get ISCO for $l = 0$ in Schwarzchild and FNS spacetimes. 
    
    \item In the CNS spacetime photon sphere and shadow are not present which implies that the central compact object is highly luminous. We studied the gravitational lensing effect in CNS spacetime where we observed that, for fixed energy photons if a charge value is much smaller than the mass of the object, the lensing of the photon is higher while a large value of charge shows a diverging trajectory.

\item
In the CNS spacetime,  we observe that
the nature of the tidal forces has some similarities with that of the RN
and differences from the Schwarzchild and FNS spacetime. 
\begin{enumerate}
 \item The Figures [{\ref{Radial}} - {\ref{Angular}}] show the behaviour of 
the radial and angular tidal forces for different values of charge $Q$ 
when the mass  $M$ is kept fixed at $1$. It is to be noted that for radial
tidal forces, there is infinite compression for CNS while for the
angular part there is infinite spaghettification, respectively.

 \item Figures ({\ref{QlessM1}} - {\ref{QgraterM1}}) show a comparison of
radial tidal force  while figures ({\ref{QlessM}} - {\ref{QgraterM}})
show the comparison for the angular tidal force for CNS, Schwarzchild, 
RN and FNS spacetimes. It is observed that both spaghettification and
compression set-in earlier in CNS as compared to the RN spacetime.

\end{enumerate}
\end{itemize}

%\newpage

\end{document}